\documentclass[11pt]{article}
\usepackage{jhep}
\usepackage[T1]{fontenc}
\usepackage{bm}
\usepackage{graphicx}
\usepackage{caption}
\usepackage{subcaption}
\usepackage[normalem]{ulem}
\usepackage{gensymb}

\title{Dynamical Axions and Gravitational Waves}
\author[a]{Djuna Croon,}
\author[b]{Rachel Houtz,}
\author[c]{Ver{\'o}nica Sanz}

\affiliation[a]{TRIUMF Theory Group, 4004 Wesbrook Mall, Vancouver, B.C. V6T2A3, Canada}
\affiliation[b]{Departamento de F{\'i}sica Te{\'o}rica and Instituto de F{\'i}sica Te{\'o}rica, IFT-UAM/CSIC,
Universidad Aut{\'o}noma de Madrid, Cantoblanco, 28049, Madrid, Spain}
\affiliation[c]{University of Sussex, Falmer Campus, Brighton, United Kingdom}
\emailAdd{dcroon@triumf.ca}
\emailAdd{rachel.houtz@uam.es}
\emailAdd{v.sanz@sussex.ac.uk}
\date{\today}
\abstract{In this paper we explore the possibility of observable gravitational waves as a manifestation of the QCD axion dynamics. In particular, we focus on dynamical axion models which solve the strong CP problem, and include the confinement of a QCD-like gauge group at the TeV scale.
We study the resulting chiral symmetry breaking phase transition for models with $N_F=3$ and $N_F=4$ light flavors using the linear sigma model. This model describes the scalar meson spectrum and its interactions, with the diagonal field $\varphi$ as the order parameter. We find that the amplitude of the gravitational wave spectrum depends on the mass of the dynamical axion $\eta'$ via the ratio $m_{\eta'}/m_\varphi$. The resulting spectra may be observed at future mid-range gravitational wave experiments such as AION/MAGIS, DECIGO, and BBO. Moreover, the TeV states can be searched for at colliders and their quantum numbers characterized, providing a unique connection between axion physics, gravitational waves and collider searches. 
}

\usepackage{natbib}
\usepackage{graphicx}

\newcommand{\beq}{\begin{equation}} 
\newcommand{\eeq}{\end{equation}}  
\newcommand{\bea}{\begin{eqnarray}}  
\newcommand{\eea}{\end{eqnarray}}  
\newcommand{\pd}[2]{\frac{\partial #1}{\partial #2}}
\newcommand{\pdd}[2]{\frac{\partial^2 #1}{\partial #2^2}}
\newcommand{\dd}[2]{\frac{d #1}{d #2}}

\allowdisplaybreaks

\definecolor{DC}{rgb}{1,0,0}

\begin{document}
\maketitle

\section{Introduction}
First order phase transitions may result in gravitational wave spectra observable at the next generation of interferometer experiments, and have therefore recently received much attention. Most of that attention has focused on the spontaneous breaking of a gauge symmetry. Particularly well-studied examples include the electroweak phase transition in Beyond the Standard Model (BSM) theories (for recent reviews, see \cite{Weir:2017wfa,Mazumdar:2018dfl}), but recently, perturbative phase transitions in hidden sectors (e.g. \cite{Jaeckel:2016jlh,Schwaller:2015tja,Croon:2018new,Croon:2018erz,Breitbach:2018ddu,Madge:2018gfl}), and phase transitions in GUT-theories \cite{Croon:2018kqn,Croon:2019kpe} have also been studied.

Here we will focus instead on chiral phase transitions, which result from the breaking of a global symmetry after a gauge group confines.
The order of such phase transitions has been the topic of decades of scientific inquiry. A famous analytic argument made by Pisarski and Wilczek \cite{Pisarski:1983ms} (PW in the following)
is based on a linear sigma model,\footnote{Note that we will also include the 't Hooft determinental interaction $\left( \mu_\Sigma\, { \rm det}\Sigma + h.c. \right) $ further on in this work, but the PW argument does not rely on its presence.}
\begin{equation} 
    V(\Sigma) = - m_\Sigma^2 {\rm Tr}\left( \Sigma \Sigma^\dagger \right) 
            + \frac\lambda2 \left[ {\rm Tr} \left( \Sigma \Sigma^\dagger \right) \right]^2
            + \frac\kappa2 { \rm Tr} \left( \Sigma \Sigma^\dagger \Sigma \Sigma^\dagger \right)\, ,
\end{equation}
where $\Sigma$ is a quark condensate.
The PW argument relies on an expansion in $\epsilon = 4 - d$, where $d$ is the number of space-time dimensions. 
To leading order in $\epsilon$, it is found that there is no infrared stable fixed point for $N_F > \sqrt{3}$. such that a first order phase transitions is expected for a large number of light flavors \cite{Pisarski:1983ms}.\footnote{This can ultimately be derived from the presence of both the $\lambda$ and the $\kappa$ coupling in the linear sigma Lagrangian.}
Of course, thermal phase transitions should be described by an effective theory in $d=3$ dimensions, as time-like fluctuations are cut off at finite temperature. The PW argument can therefore be taken as a guide only. Nevertheless, the result that chiral phase transitions with $N_F\geq 3$ are first order is commonly accepted, and demonstrated on the lattice for $N_F = 6$ \cite{Iwasaki:1995ij}.

The gravitational wave phenomenology of first order confining phase transitions has implications for pressing questions beyond the Standard Model (SM). One such open question is whether or not QCD violates CP, known as the strong CP problem.
The experimentally small CP violating coupling, $\bar\theta = \theta_{QCD} + \text{arg } \text{det}\,M  < 10^{-10}$~\cite{Baker:2006ts, Engel:2013lsa} is unexplained in the SM.
The most common dynamical explanation for the lack of CP violation in the strong sector is the introduction of a global $U(1)$ Peccei-Quinn (PQ) symmetry that allows the CP violating parameter to be rotated to zero via field redefinitions~\cite{Peccei:1977hh}. Upon spontaneous breaking of the $U(1)_{PQ}$, the resulting associated pseudo-Goldstone boson that couples anomalously to the field strength of QCD is the axion~\cite{Wilczek:1977pj, Weinberg:1977ma}.

The strong CP problem can also be solved using massless quarks~\cite{Kaplan:1986ru}. In such models, a $U(1)$ chiral rotation of the massless quark fields rotates $\bar\theta$ to zero~\cite{tHooft:1976snw}. This $U(1)$ is also spontaneously broken by the chiral condensation when the quarks' color group confines. The resulting pseudo-Goldstone boson that couples anomalously to the field strength of the confining group is the $\eta'$, composed of massless quarks. In this class of models, the $\eta'$ plays the role of the composite, or dynamical, axion. 

The behavior of the $\eta'$ can be studied in the framework of new exotic confining color groups, which are prevalent in model building to address the strong CP problem~\cite{ Kim:1984pt,Choi:1985cb,  Rubakov:1997vp,Tye:1981zy, Berezhiani:2000gh, Hook:2014cda, Fukuda:2015ana,Draper:2016fsr, Gherghetta:2016fhp, Dimopoulos:2016lvn, Gaillard:2018xgk, Gavela:2018paw}. Heavy  or visible axion models predominately utilize exotic color groups to alter the typical $(m_a, f_a)$ relationship~\cite{Tye:1981zy, Rubakov:1997vp, Berezhiani:2000gh,Hook:2014cda, Fukuda:2015ana, Gherghetta:2016fhp, Dimopoulos:2016lvn, Agrawal:2017evu, Agrawal:2017ksf,  Gaillard:2018xgk, Gavela:2018paw}.
There are generic features of chiral phase transitions in dynamical axion models that set them apart from general confining hidden sector models. In particular, axion models that utilize an exotic color group typically have matter charged under both QCD and the exotic group. If any of that matter is fermionic, the three colors of QCD guarantee at least an approximate $N_F=3$ flavor symmetry from the point of view of the exotic group, and this is exactly the minimum flavor symmetry required for a first order phase transition. Also, the pions associated with the broken $SU(N_F)$ symmetry will get quadratically divergent mass terms due to interactions with QCD, and so such models have a generic form for the explicit symmetry breaking from QCD in the linear sigma model. 
Although one could imagine a number of ways to achieve the dynamical axion solution, in this paper we often use the example discussed in~\cite{Gaillard:2018xgk} as a benchmark. 

Gravitational wave signals from confining phase transitions have recently been studied in dark QCD-like models with $N_F =3$ \cite{Tsumura:2017knk,Bai:2018dxf,Helmboldt:2019pan}.
In this paper, we will focus on the relation between the dynamical axion and the gravitational wave signature of the chiral phase transition, for $N_F = 3$ and $N_F = 4$.
Using the linear sigma model as the low-energy effective theory, we find that the gravitational wave predictions depend sensitively on the mass of the dynamical axion.\footnote{An alternative choice is to study gravitational waves from the chiral phase transition using the Nambu-Jona-Lasinio model \cite{Aoki:2017aws,Helmboldt:2019pan}.} This is an interesting result, which invites further study of the phenomenology of these models.

\section{The dynamical axion in the linear sigma model}
\label{sec:axion-linear-sigma}

To discuss the origin of the dynamical heavy axion we consider a typical framework, an exotic $SU(\tilde N)$ color sector that confines at a scale $\Lambda \gg \Lambda_{QCD}$, connected to the SM through a quark field $\psi$ charged under QCD and $SU(\tilde N)$, shown in Table~\ref{tab:nf=3}. This quark field guarantees an approximate $N_F\geq 3$ flavor symmetry upon $SU(\tilde N)$ confinement as $\psi$ is a triplet of QCD. 

\begin{table}[h]
\[
\begin{array}{c|cc}
    & SU(3)_{QCD} \ \ \    & SU(\tilde N) \\
\hline
\psi&   \Box        & \Box
\end{array}
\]
\caption{ The messenger field between the SM QCD and the $SU(\tilde N)$ confining exotic group. }
\label{tab:nf=3}
\end{table}

Here we study the chiral symmetry breaking phase transition associated with the $SU(\tilde N)$ confinement.  Upon confinement, the chiral $SU(N_F)_L \times SU(N_F)_R $ symmetry is broken to $SU(N_F)_V$, where $N_F$ is the number of flavors in the exotic sector.  Below confinement, the dynamics of the hidden sector are modeled by the linear sigma model. The resulting pion fields we discuss are bound states made out of exotic quarks. 

If $\psi$ is massless, this field is important to the solution of the strong CP problem. Chiral rotations on the $\psi$ field can be used to rotate away the CP violating angle of either QCD or $SU(\tilde N)$, though not both independently.\footnote{Additional model building is needed to solve the strong CP problem completely in these types of models. For example, either another massless quark field is introduced or a symmetry relates $\theta_{QCD}$ and $\tilde \theta$.} Moreover, a vanishing tree-level mass for $\psi$ ensures that $\psi$ is active when $SU(\tilde N)$ confines, protecting the first order nature of the chiral phase transition. Below confinement, $\psi$ will be hidden in the resulting bound states. One of those bound states, the $\eta'$, will be the pseudo-Goldstone boson of $U(1)_A$, which is broken explicitly by the instantons of $SU(\tilde N)$. This particle will then couple anomalously to $G\tilde G$, and plays the role of the dynamical axion in the theory below the confinement scale.

\subsection{Low energy effective theory}
At low energies, the dynamics of the quark condensate $\Sigma_{ij} \sim \langle {\bar \psi_{Rj}} {\psi_{Li}} \rangle $ can be described by a linear sigma model.
The effective potential for the dynamical field $\Sigma$ which transforms as a $(\Box,\bar{\Box})$ under $SU(N_F)_L \times SU(N_F)_R$ is given by
\begin{equation} \label{eq:linearsigma}
    V(\Sigma) = - m_\Sigma^2 {\rm Tr}\left( \Sigma \Sigma^\dagger \right) 
            -\left( \mu_\Sigma\, { \rm det}\Sigma + h.c. \right)
            + \frac\lambda2 \left[ {\rm Tr} \left( \Sigma \Sigma^\dagger \right) \right]^2
            + \frac\kappa2 { \rm Tr} \left( \Sigma \Sigma^\dagger \Sigma \Sigma^\dagger \right) \ .
\end{equation}
The chiral condensate $\langle \Sigma_{ij} \rangle \sim f_\Sigma \delta_{ij}$ spontaneously breaks the global chiral symmetry $SU(N_F)_L\times SU(N_F)_R \to SU(N_F)_V$. This effect is captured by the linear sigma model when:
\begin{align}
-  \frac{  m_\Sigma^4 }{  \kappa + N_F \lambda} < 0 \ .
\end{align}
$\Sigma$ can be decomposed as:
\begin{equation}
\Sigma_{ij}
    = \frac{ \varphi + i \eta' }{ \sqrt{ 2 N_F} } \delta_{ij} 
        + X^a T^a_{ij}
        + i \pi^a T^a_{ij} \ ,
\label{eq:sigma-decomposition}
\end{equation}
where  $T^a_{ij}$ are the generators of the $SU(N_F)_{L, R}$ symmetry. The $\pi^a$ are the pseudo-Goldstone bosons associated with the broken combination $SU(N_F)_A$. The $\varphi$ and $X^a$ fields are massive bound states associated with the preserved $SU(N_F)_V \times U(1)_V$ symmetry.

In the $\mu_\Sigma \to 0$ limit, the chiral symmetry of $V(\Sigma)$ is enhanced to $U(N_F)_L \times U(N_F)_R$, which contains an extra  $U(1)_A$ restored in~\eqref{eq:linearsigma}, but spontaneously broken by the chiral condensate. The $\eta'$ in~\eqref{eq:sigma-decomposition} is the Goldstone boson associated with the spontaneous $U(1)_A$ breaking. The $\mu_\Sigma\to0$ limit does not properly describe properties of a confining gauge group, as we know the axial anomaly \it explicitly \rm breaks $U(1)_A$ to $Z_{N_F}$ by quantum effects. These quantum effects originate from instantons of the confining $SU(\tilde N)$. Indeed, it was recognized in \cite{Pisarski:1983ms} that the sum of instanton and anti-instantons generate $2 N_F$-point interactions of the form ${ \rm det}\Sigma + h.c.$. Thus the $\mu_\Sigma$-term in the linear sigma model captures the $U(1)_A$-breaking instanton effects in the low energy theory. However, identifying $\mu_\Sigma$ with the temperature-dependent instanton density \cite{Pisarski:1980md}, one can draw the conclusion that $\mu_\Sigma (T)$ vanishes for $T\rightarrow \infty$, as Debye screening shields electric field fluctuations at high temperature. We will discuss this issue in more detail in section \ref{sec:instantons}.

Crucially, the $\eta'$ is anomalously coupled to $G\tilde G$, making it a dynamical axion. Above the confinement scale, the strong CP problem is solved by the presence of the massless $\psi$ quarks. The $\eta'$ is a bound state composed of $\psi$ quarks, and therefore its connection to the global $U(1)$ that rotates the CP violating phase is explicit. Note that in the presence of a mass term for the $\psi$ quarks, the $U(1)_A$ is classically explicitly broken and no longer meets the criteria for a dynamical axion.\footnote{Explicit mass terms for the $\psi$ quark would lead to an addition of a $M_q \rm{Tr} \Sigma$ term in~\eqref{eq:linearsigma}. If this term is present, then rotations like $\Sigma \to e^{i\phi} \Sigma$ would not be able to remove complex phases from both the $\mu_\Sigma$ and $M_q$ simultaneously~\cite{tHooft:1986ooh}. A complex phase in $M_q$ is evidence of possible CP violation in the strong sector.
}

The $\eta'$ axion obtains a mass due precisely to the explicit $U(1)_A$-breaking effects from the instantons of the confining group. This explicit breaking should result in an axion potential, which typically takes the form:
\begin{align} 
\mathcal{L}_{ \rm s.b.}
    \ni \Lambda^4 \cos\left( \frac{\eta'}{f_a}\right) \ .
\label{eq:axion-potential}
\end{align}
Using~\eqref{eq:axion-potential} we can connect the $\mu_\Sigma$ parameter in the linear sigma model to the physical axion mass predicted by the axion's couplings to $G \tilde G$, motivating our choices for linear sigma model parameters in our gravitational wave signal analysis in Section~\ref{sec:GW-spectra}. 

The QCD-colored pions resulting from the spontaneous symmetry breaking should also receive masses since QCD explicitly breaks $SU(N_F)$. The QCD-induced mass of the pions can be included in the linear sigma model by adding the explicit flavor symmetry breaking term
\begin{align}
V(\Sigma)
    &\ni \xi \left( 
    \text{Tr} Q^a \Sigma \Sigma^\dagger {Q^a}^\dagger
    -\text{Tr} Q^a \Sigma {Q^a}^\dagger \Sigma^\dagger 
    - \Sigma \text{Tr} Q^a\Sigma^\dagger{Q^a}^\dagger
    + \text{Tr} \Sigma Q^a {Q^a}^\dagger\Sigma^\dagger\right) \ .
\end{align}
This potential term is motivated by the fact that the QCD flavor breaking comes from the kinetic term $D^\mu \Sigma \left( D_\mu \Sigma \right)^\dagger \supset g^2 G_a^\mu G^a_\mu \rm{Tr}\left[ \left( Q_L^a \Sigma - \Sigma Q_R^a \right) \left(  \Sigma^\dagger {Q_L^a}^\dagger - {Q_R^a}^\dagger \Sigma^\dagger \right) \right]$, where $D_\mu \Sigma = \partial_\mu \Sigma - i g G_\mu Q_L \Sigma + i g G_\mu \Sigma Q_R$.
The form of $Q_{L, R}^a$ will depend on the number of flavors and how QCD interacts with the fields inside $\Sigma$. This is worked out explicitly for $N_F = 3$ and $N_F=4$ below. 

Evidence of QCD breaking the flavor symmetry comes in the form of quadratically divergent gluon loops driving the masses of the pions up towards $\Lambda$. Their zero-temperature loop-induced masses are:
\begin{align}
m^2(\pi_R) &\approx 3 C_2(R) \frac{ \alpha_c }{ 4 \pi } \Lambda^2 \ .
\label{eq:pion-mass}
\end{align}
$C_2(R)$ is the quadratic Casimir of $R$, where the pion is in the $R$ representation of $SU(3)_{QCD}$.  In general, if the dynamical axion solves the QCD strong CP problem, the massless quarks that compose the axion should be coupled to QCD. Typically, the massless messenger quark present at $SU(\tilde N)$ confinement will leave behind some pion states charged under QCD. Thus, pion masses of the form~\eqref{eq:pion-mass} are a generic feature of dynamical axion models with hidden chiral phase transitions.

Octet pions are present in both the $N_F=3$ and $N_F=4$ models, while the triplets are only present in the $N_F=4$ model, as will be discussed in Section~\ref{sec:mass-nf4}. Colored states can be searched for at colliders via their gluon couplings.
Collider searches provide a lower limit on the masses of the octet and triplet pions~\cite{GoncalvesNetto:2012nt, Degrande:2014sta, Aaboud:2016uth},  
\begin{align}
&m (\pi_8) \gtrsim 770 \text{ GeV} 
    \label{eq:octet-bound}
&& m (\pi_3) \gtrsim 890 \text{ GeV} \ .
\end{align}
Using ~(\ref{eq:octet-mass}) and (\ref{eq:triplet-mass}), the lower limits on pion masses provide a lower limit for the on the confinement scale:
\begin{align}
    \Lambda \gtrsim 2.9 \text{ TeV} \ . 
\end{align}

The other relevant bound comes from the lightest dynamical axion. Here the dynamical scale for the axion is the chiral symmetry breaking scale $f_a = f_\Sigma$, where $4\pi f_\Sigma \geq \Lambda$. For a confinement scale $\mathcal{O}(\text{TeV})$, a lower bound on the light axion mass comes from beam dump experiments, and an axion heavier than $100$ MeV easily avoids these bounds. The colliders LEP, CDF, and LHC, probe but do not fully cover axions with a dynamical scale $\mathcal{O}(\text{TeV})$ and mass $\mathcal{O}(\text{TeV})$~\cite{Mimasu:2014nea,Jaeckel:2015jla,Brivio:2017ije}.

\subsection{Meson masses}\label{sec:masses}

In this section we compute the spectrum of meson states, required to compute the one-loop thermal corrections to the effective potential discussed in the next section.
\subsubsection{$N_F =3$}

The $N_F = 3$ case is representative of  models with a massless quark field charged as bifundamental under $SU(3)_{QCD}$ and a confining $SU(\tilde N)$, as shown in Table~\ref{tab:nf=3}. Models in Ref.~\cite{Hook:2014cda} and~\cite{Gaillard:2018xgk} have a confining exotic color (gauge) group with an $SU(3)_L \times SU(3)_R$ (global) chiral symmetry. The $SU(3)$ flavor symmetry is explicitly broken by QCD. The effects of this are captured by adding the $\xi$-term to the linear sigma model:
\begin{align}
V_{\xi}( \Sigma ) 
	&\supset  \xi \left[ \text{Tr} \left(  T^a \Sigma  \Sigma^\dagger T^a \right)
		-\text{Tr} \left(  T^a \Sigma T^a \Sigma^\dagger \right)
		-\text{Tr} \left(  \Sigma T^a \Sigma^\dagger T^a\right)
		+\text{Tr} \left(  \Sigma T^a T^a \Sigma^\dagger \right)
		\right] \ ,
		\label{eq:VT0NF3}
\end{align}
where $T^a$ are the $SU(3)$ generators. We find upon minimizing the potential that
\bea \label{eq:vevNf3} f_\Sigma = \frac{\sqrt{\frac{3}{2}} \left(\mu_\Sigma +\sqrt{\mu_\Sigma ^2+4 m_\Sigma ^2 (\kappa +3 \lambda )}\right)}{\kappa +3 \lambda }\ .\eea 
The masses of the $\eta'$ and the pion are
\begin{align}
m_{\eta'}^2
	&=  \frac{ 3 \mu_\Sigma }{ 2( \kappa + 3 \lambda)  } \left( \mu_\Sigma - \sqrt{ 4 m_\Sigma^2 (\kappa + 3\lambda) + \mu_\Sigma^2 } \right) \ , 
\!\!\!\!\! \!\!\!\!\! \!\!\!\!\! \!\!\!\!\! \!\!\!\!\! \!\!\!\!\! 
&&m^2_{\pi^a}
	= 3\xi \ . 
	\label{eq:massesNf3}
\end{align}
These fields are pseudo-Goldstone bosons associated with the $SU(N_F)_A \times U(1)_A$ symmetries spontaneously broken by the chiral condensate. The $\eta'$ gets its mass from the anomalous $U(1)_A$ breaking. In the linear sigma model, the $\mu_\Sigma$-term breaks $U(1)_A$ and so $m_{\eta'}$ should be proportional to $\mu_\Sigma$. 

The octet pions get their mass from interactions with QCD, and so their mass should be proportional to $\xi$. Given that we know $\xi$ is generated by a quadratically divergent effect, we can use~(\ref{eq:pion-mass}) to estimate $\xi \sim 3 \frac{\alpha}{4\pi} \Lambda^2 $.

The masses of the heavy bound states are
\begin{align}
m_\varphi^2	
	&= 2 m_\Sigma^2 + \frac12 \mu_\Sigma \ \frac{ \mu_\Sigma + \sqrt{ 4 m_\Sigma^2\left(  \kappa + 3 \lambda  \right) + \mu_\Sigma^2 } }{  \kappa + 3 \lambda }
\notag \\
m^2_{X^a}
    &= \frac{ 2 m_\Sigma^2 \left( \kappa + 3 \lambda \right) + \mu_\Sigma \left( 2 \kappa + 3 \lambda \right) \left( \mu_\Sigma + \sqrt{ 4 m_\Sigma^2 \left( \kappa + 3\lambda \right) + \mu_\Sigma^2 } \right) }{ \left( \kappa + 3\lambda \right)^2 }
    + 3\xi  \ .
\end{align}
The heavy states have masses proportional to $m_\Sigma^2$. They are held together by the binding energy from the confining gauge group and should have masses near the confinement scale $m_\Sigma^2 \sim \Lambda^2$.

\subsubsection{$N_F =4$}
\label{sec:mass-nf4}

\begin{table}[h!]
\[
\begin{array}{c|cc}
    & SU(3)_{QCD}   & SU(\tilde N) \\
\hline
\psi&   \Box        & \Box  \\
\chi&   1           & \Box
\end{array}
\]
\caption{The massless quark content that gives an approximate $SU(4)_L \times SU(4)_R$ chiral symmetry broken by the chiral condensate. Interactions with $QCD$ explicitly break the $SU(4)$ flavor symmetry. }
\label{tab:nf=4}
\end{table}

The $N_F = 4$ case we examine is given in Table~\ref{tab:nf=4}. This is representative of models in which the exotic confining group requires a second independent massless field to independently rotate away the $\theta$-angle of the confining $SU(\tilde N)$ group. Examples of axion models that have this flavor structure near exotic confinement are Ref.~\cite{Choi:1985cb} and Model I of Ref.~\cite{Gaillard:2018xgk}.

QCD explicitly breaks the $SU(4)$ flavor symmetry. The effects of this are captured in the addition of the $\xi$-term to the linear sigma model:
\begin{align}
V_{\xi}( \Sigma ) 
	&=  \xi \left( \text{Tr}  Q_L^a \Sigma  \Sigma^\dagger Q_L^a
		-\text{Tr}  Q_L^a \Sigma Q_R^a \Sigma^\dagger 
		-\text{Tr}  \Sigma Q_R^a \Sigma^\dagger Q_L^a
		+\text{Tr}  \Sigma Q_R^a Q_R^a \Sigma^\dagger 
		\right)
\label{eq:VT0NF4}
\end{align}
where $Q_L^a = Q_R^a \equiv Q^a$ is a $4\times4$ matrix with the top left $3\times3$ submatrix given by $SU(3)_{QCD}$ generators:
\begin{align}
Q^a 	&= \left( \begin{array}{cc}
		T^a	& 0	\\
		0	& 0
		\end{array} \right)
\end{align}
and the fourth row and column filled with zeroes. $\Sigma_{ij} \sim \langle {\bar \Psi}_{Rj} \Psi_{Li} \rangle$, where the flavor multiplet  $\Psi_i = \left( \psi_i, \chi \right)$ contains three $\psi_i$ quarks and one $\chi$. Since only $\psi$ is charged under QCD, the explicit symmetry breaking effects of the $\xi$-term should only effect the bound states containing $\psi$ quarks.

We know that when $SU(4)\to SU(3)\times U(1)$, fields in the Adjoint representation of $SU(4)$ break as
\begin{equation}
15 \to 1 + 3 + \bar3 + 8
\end{equation}
where the $15$ is representation of the $\pi^a$ of SU(4), the $1$ field is $\eta_\chi'$, and the $3, \bar3, 8$ fields are the QCD-charged pions. Even when the $SU(4)\to SU(3)$ breaking does not happen, this breaking pattern gives us a hint as to which representations of $SU(3)$ are living inside a representation of $SU(4)$. Given~\eqref{eq:pion-mass}, we expect that $\pi^a$ contains mass eigenstates
\begin{align} 
m^2(\pi_8) &\approx \frac{ 9 \alpha_c }{ 4 \pi } \Lambda^2
\label{eq:octet-mass}
\\
m^2(\pi_3) &\approx \frac{\alpha_c }{ \pi } \Lambda^2 \ ,
\label{eq:triplet-mass} 
\end{align}
where $\pi_8$ are the color octet pions and $\pi_3$ are the color triplet pions.

We find upon minimizing the potential that
\begin{align}
f_\Sigma^2 &= \frac{ 8 m_\Sigma^2 }{ \kappa + 4 \lambda - \mu_\Sigma} \ \ .
\end{align}
The $\eta'$ and the pion masses are
\begin{align}
m_{\eta'_\psi}^2
	&= \frac{ 4 m_\Sigma^2 \mu_\Sigma }{ \kappa + 4 \lambda - \mu_\Sigma } \ , 
\!\!\!\!\! \!\!\!\!\! \!\!\!\!\! \!\!\!\!\! \!\!\!\!\! \!\!\!\!\! \!\!\!\!\! \!\!\!\!\! \!\!\!\!\! \!\!\!\!\! \!\!\!\!\!
&&m_{\eta'_\chi}^2	
	    = 0 
\\
m_{\pi_8}^2	
    &= 3\xi
 \ , 
\!\!\!\!\! \!\!\!\!\! \!\!\!\!\! \!\!\!\!\! \!\!\!\!\! \!\!\!\!\! \!\!\!\!\! \!\!\!\!\! \!\!\!\!\! \!\!\!\!\! \!\!\!\!\! 
&&m_{\pi_3}^2	
    = \frac43\xi \ \ .
\end{align}
The pions $\pi_3$ and $\pi_8$ have masses proportional to $\xi$. The mass of the QCD-colored pions are given by the QCD's quadratically divergent contribution to the pion self energy,~(\ref{eq:octet-mass}) and~(\ref{eq:triplet-mass}). Using this, we can estimate $
\xi \sim \frac{ 3 \alpha_c }{ 4\pi } \Lambda^2 $.

The $\eta_\chi'$ and $\eta_\psi'$ are the singlet pseudoscalars that couple to the confining group's $G \tilde G$. The confinement dynamics that break $U(1)_A$, namely, the instanton effects, are described by the $\mu_\Sigma$-term of the linear sigma model. Since both $\eta_\chi'$ and $\eta_\psi'$ couple identically to $G \tilde G$, these instantons can only give mass to one eigenstate. This explains the spectrum. The $\eta_\psi'$ mass is proportional to $\mu_\Sigma$ since its mass corresponds to explicit $U(1)_A$ breaking. Then the linear sigma model predicts the light eigenstate $m_{\eta_\chi'}=0$. The $\eta_\chi'$ is not exactly massless, but obtains its mass via mixing with the SM pion below QCD confinement.

The mass of the heavy bound states associated with the unbroken generators, $X^a$ and $\varphi$ are:
\begin{align}
m_\varphi^2
    &= 2 m_\Sigma^2\ \ , 
\!\!\!\!\! \!\!\!\!\! \!\!\!\!\! \!\!\!\!\! \!\!\!\!\! \!\!\!\!\! \!\!\!\!\! 
&&m_{X_1}^2
	= 2 m_\Sigma^2  \frac{  \kappa + \mu_\Sigma  }{ \kappa + 4 \lambda - \mu_\Sigma } 
\\
m_{X_8}^2	
    & =  2 m_\Sigma^2 \frac{  \kappa + \mu_\Sigma }{ \kappa + 4 \lambda - \mu_\Sigma } + 3 \xi \ \ , 
\!\!\!\!\! \!\!\!\!\! \!\!\!\!\! \!\!\!\!\! \!\!\!\!\! 
&&m_{X_3}	
	   =  2 m_\Sigma^2 \frac{  \kappa + \mu_\Sigma }{ \kappa + 4 \lambda - \mu_\Sigma } + \frac43 \xi
\end{align} 
Because the 15 $X_R^a$ are no longer degenerate, the mass eigenstates have been renamed $X_1, X_3, X_8$ for the particles charged as singlets, triplets, and octets under QCD, respectively. All the heavy states have masses proportional to $ m_\Sigma^2$. They are held together by the binding energy from the confining gauge group and should have masses near the confinement scale $m_\Sigma^2 \sim \Lambda^2$. 

The light state $\eta_\chi'$ will be present at QCD confinement and will couple anomalously to QCD's field strength tensor through  triangle diagram involving $\psi$ quarks. Thus, the $\eta_\chi'$ is also a dynamical axion, and without any additional mass sources will mix with the SM pions and yield an invisible axion with the typical relationship $m_a f_a \sim m_\pi f_\pi$ characteristic of invisible axion models~\cite{ Kim:1979if,Shifman:1979if, Dine:1981rt, Zhitnitsky:1980tq, Choi:1986zw}.

Ref.~\cite{Gaillard:2018xgk} provides an example model where additional mass contributions to the axion potential raise the $\eta_\chi'$ mass. This model is thus a heavy axion model with an $N_F=4$ flavor symmetry at $SU(\tilde N)$ confinement.

The new mass contributions come from small-sized instantons in the UV theory that couple only to $\chi$ and not to $\psi$. In this analysis, the effect is captured by adding a $\mu_{SSI}$-term to the linear sigma model:
\begin{align}
    V(\Sigma) 
        &\supset \mu_{SSI} \rm{Tr} \left( P_\chi \Sigma P_\chi \Sigma^\dagger P_\chi \right) \ ,
\end{align}
where
\begin{align}
    P_\chi
        &= \left( \begin{array}{cc}
            0_{3\times3}    & 0 \\
            0               & 1 
            \end{array}\right)
\end{align}
picks out only the $\chi$ component of $\Psi_i = \left( \psi_i, \chi \right)$. This $\mu_{SSI}$-term should account for mass contributions to any state that contains a $\chi$ quark, and so will affect both $\eta_\psi'$ and $\eta_\chi'$,\footnote{Despite their names, $\eta_\chi'$ and $\eta_\psi'$ are both combinations of $\psi$ and $\chi$ quarks. The $\eta_\psi'$ is associated with $1_{4\times4}$ and the $\eta_\chi'$ is associated with the $T^{(15)}$ generator of $SU(4)$.} raising the lightest mass eigenstate. 

Including this new $\mu_{SSI}$-term, we find upon minimizing the potential that
\begin{align}
f_\Sigma^2 &= \frac{ 8 m_\Sigma^2 - 2 \mu_{SSI} }{ \kappa + 4 \lambda - \mu_\Sigma} \ \ .
\end{align}
The $\eta'$ and the pion masses are
\begin{align}
m_{\eta'_\psi}^2
	&= \frac{ \left(  4 m_\Sigma^2 - \mu_{SSI} \right) \mu_\Sigma }{ \kappa + 4 \lambda - \mu_\Sigma } \ , 
\!\!\!\!\! \!\!\!\!\! \!\!\!\!\! \!\!\!\!\! \!\!\!\!\! \!\!\!\!\! \!\!\!\!\! \!\!\!\!\! \!\!\!\!\! \!\!\!\!\!
&&m_{\eta'_\chi}^2	
	    = \frac12 \mu_{SSI}
\\
m_{\pi_8}^2	
    &= 3\xi + \frac14 \mu_{SSI}
 \ , 
\!\!\!\!\! \!\!\!\!\! \!\!\!\!\! \!\!\!\!\! \!\!\!\!\! \!\!\!\!\! \!\!\!\!\! \!\!\!\!\! \!\!\!\!\! \!\!\!\!\! \!\!\!\!\! 
&&m_{\pi_3}^2	
    = \frac43\xi + \frac14 \mu_{SSI} \ \ .
\end{align}
Note that now both mass eigenstates contain $\mu_{SSI}$ contributions, and so depending on the strength of the SSI instantons interacting with the $\chi$ field, the lightest dynamical axion can be made heavy. 

The masses of the heavy bound states associated with the unbroken generators, $X$ and $\varphi$ are now
\begin{align}
m_\varphi^2
    &= 2 m_\Sigma^2 - \frac12 \mu_{SSI}
\\
m_{X_1}^2
	&=   \frac{ 2 m_\Sigma^2 \left( \kappa + \mu_\Sigma \right) + \mu_{SSI} \left( 2 \lambda - \mu_\Sigma \right) }{ \kappa + 4 \lambda - \mu_\Sigma } 
\\
m_{X_8}^2	
    & =  \frac{   2 m_\Sigma^2 \left( \kappa + \mu_\Sigma \right)  - \frac14 \mu_{SSI} \left(  3\kappa + 4\lambda + \mu_\Sigma \right)  }{ \kappa + 4 \lambda - \mu_\Sigma }  + 3 \xi  
\\
m_{X_3}	
	  &=   \frac{ 2 m_\Sigma^2 \left( \kappa + \mu_\Sigma \right) - \frac14 \mu_{SSI} \left(  3\kappa + 4\lambda + \mu_\Sigma \right) }{ \kappa + 4 \lambda - \mu_\Sigma } + \frac43 \xi \ ,
\end{align} 
where the $\mu_{SSI}$ provides an additional mass source, though all heavy bound states still contain $m_\Sigma^2\sim\Lambda^2$ contributions as well.

\section{Gravitational waves from a chiral phase transition}
In this section we study the finite-temperature one-loop effective potential generated by the scalar mesons, and explore the parameter space leading to a first-order phase transition. The gravitational spectrum of the phase transition is then derived for two distinct cases, which differ in the number of light fermions: $N_F= 3$ and $N_F =4$.  

\subsection{One loop effective potential at finite temperature}\label{sec:thermalV}

\begin{figure}
    \centering
    \includegraphics[width=0.55\textwidth]{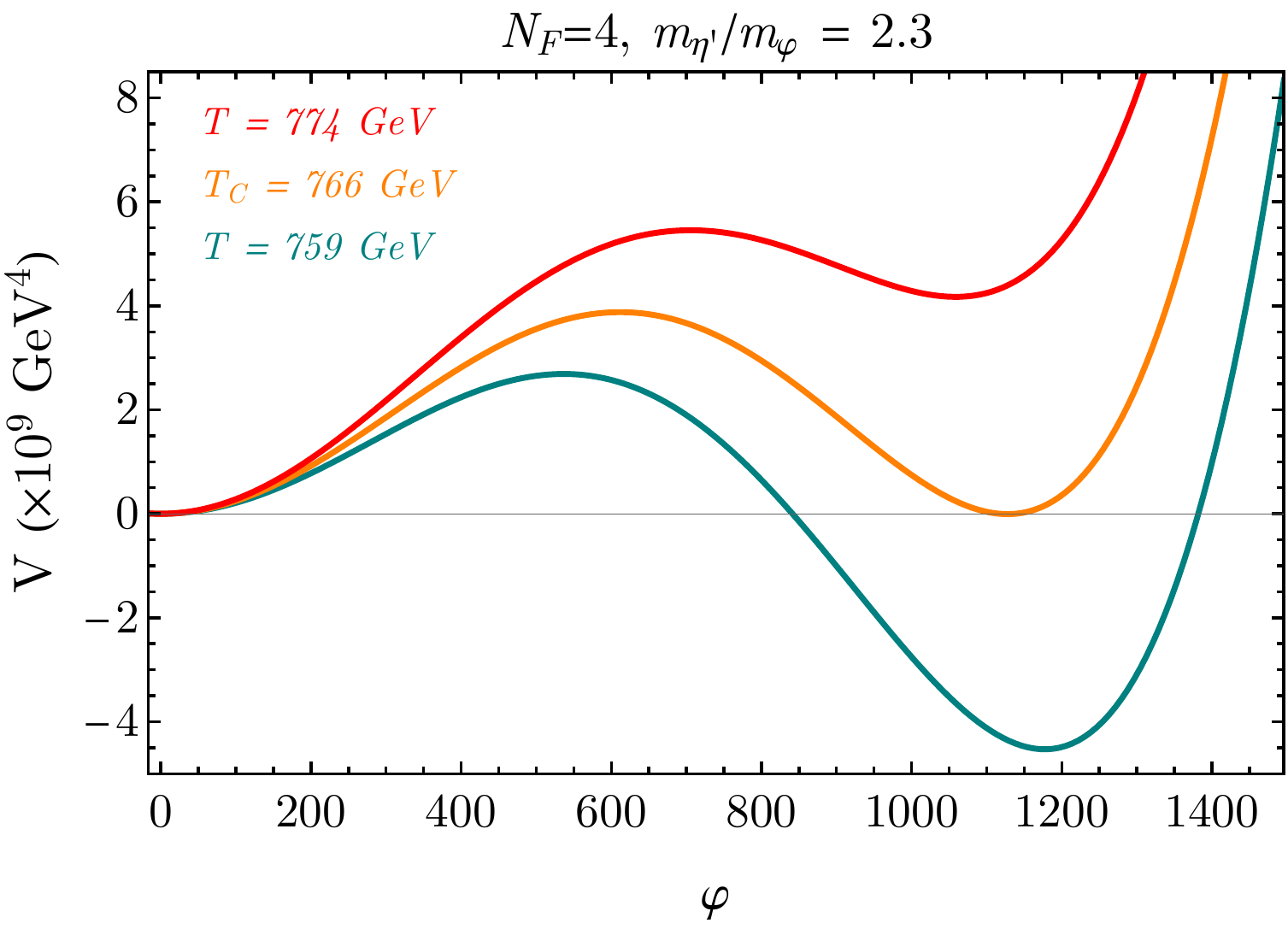}
    \caption{Thermal potential for a benchmark with $N_F=4$. At zero temperature, the barrier disappears, while at high temperature, the global vacuum has vanishing vacuum expectation value $\langle \varphi \rangle = 0$.}
    \label{fig:thermalpotential}
\end{figure}

At one loop, we consider the following daisy-resummed thermal corrections to the potentials \eqref{eq:linearsigma}, \eqref{eq:VT0NF3} and \eqref{eq:VT0NF4},\footnote{An alternative (non-perturbative) approach is the dimensionally reduced effective theory \cite{Niemi:2018asa,Gould:2019qek}.}
\bea
    V(\Sigma,T) &=& V(\Sigma) + V_\chi(\Sigma)+  V_{T \neq 0},\\
    V_{T \neq 0} &=&  \sum _{i} \frac{T^4}{2 \pi ^2} n_i J_B\left( \frac{m_i^2+\Pi _i }{T^2}  \right),  \\
J_B(m^2) & = &  \int_0^{\infty} dx\, x^2 \log \left(1-e^{-\sqrt{x^2+m^2}}\right) .
\eea
Here $\Sigma$ is the linear sigma field \eqref{eq:sigma-decomposition}. We consider the zero temperature (Coleman-Weinberg) contribution to the potential to be a redefinition of the parameters in our zero temperature potential \eqref{eq:linearsigma}, and do not add it explicitly. The chiral phase transition describes the condensation of the diagonal combination $\varphi$. Therefore, the relevant thermal corrections are by all scalar states that couple to $\varphi$:  $m_i$ are their thermal, field-dependent masses and $n_i$ are their multiplicities. 
For $N_F=3$, the field dependent thermal masses $m_i$ are given by,
\bea \notag m^2_\varphi + \Pi_\varphi &=&
\frac{1}{6} \left(\varphi  \left(3 \kappa  \varphi +9 \lambda  \varphi -2 \sqrt{6} \mu_\Sigma \right)-6 m_\Sigma^2+T^2 (3 \kappa +5 \lambda )\right) \\ \notag
m^2_{\eta'} + \Pi_{\eta'} &=& \frac{1}{6} \left(\varphi  \left(\kappa  \varphi +3 \lambda  \varphi +2 \sqrt{6} \mu_\Sigma \right)-6 m_\Sigma^2+T^2 (3 \kappa +5 \lambda )\right) \\ \notag
m^2_{X} + \Pi_{X}&=& \frac{1}{6} \left(3 \kappa  \varphi ^2+3 \lambda  \varphi ^2+\sqrt{6} \mu_\Sigma  \varphi -6 m_\Sigma^2-18 \xi +T^2 (3 \kappa +5 \lambda )\right)\\ 
m^2_{\pi} + \Pi_{\pi}&=& \frac{1}{6} \left(\kappa  \varphi ^2+3 \lambda  \varphi ^2-\sqrt{6} \mu_\Sigma  \varphi -6 m_\Sigma^2-18 \xi +T^2 (3 \kappa +5 \lambda )\right),
\eea
(as also found in \cite{Bai:2018dxf}), and for $N_F = 4$, they are given by
\bea \notag m^2_\varphi + \Pi_\varphi &=& \frac{1}{24} \left(9 \kappa  \varphi ^2+36 \lambda  \varphi ^2-9 \mu_\Sigma  \varphi ^2+6 \mu_\text{SSI}-24 m_\Sigma^2+2 T^2 (8 \kappa +17 \lambda )\right)
 \\ \notag
m^2_{\eta'}+ \Pi_{\eta'} &=& \frac{1}{8} \left(\varphi ^2 (\kappa +4 \lambda +3 \mu_\Sigma )+2 \mu_\text{SSI}\right)-m_\Sigma^2+\frac{1}{12} T^2 (8 \kappa +17 \lambda ) \\ \notag
m^2_{X_8} + \Pi_{X_8} &=& \frac{1}{8} \varphi ^2 (3 \kappa +4 \lambda +\mu_\Sigma )-m_\Sigma^2-3 \xi +\frac{1}{12} T^2 (8 \kappa +17 \lambda ) 
\\
\notag
m^2_{X_3} + \Pi_{X_3}&=& \frac{1}{24} \left(9 \kappa  \varphi ^2+12 \lambda  \varphi ^2+3 \mu_\Sigma  \varphi ^2-24 m_\Sigma^2-32 \xi +2 T^2 (8 \kappa +17 \lambda )\right) 
\\ \notag
m^2_{\pi_8} + \Pi_{\pi_8}&=& \frac{1}{8} \varphi ^2 (\kappa +4 \lambda -\mu_\Sigma )-m_\Sigma^2-3 \xi +\frac{1}{12} T^2 (8 \kappa +17 \lambda ) 
\\  \notag
m^2_{\pi_3} + \Pi_{\pi_3} &=& \frac{1}{24} \left(3 \kappa  \varphi ^2+12 \lambda  \varphi ^2-3 \mu_\Sigma  \varphi ^2-24 m_\Sigma^2-32 \xi +2 T^2 (8 \kappa +17 \lambda )\right) \\ \notag
m^2_{\eta_{\psi}'} + \Pi_{\eta'_{\psi}}&=& \frac{1}{24} \left(3 \varphi ^2 (3 \kappa +4 \lambda +\mu_\Sigma )+18 \mu_\text{SSI}-24 m_\Sigma^2+2 T^2 (8 \kappa +17 \lambda )\right)\\
m^2_{\eta'_{\chi}} + \Pi_{\eta'_{\chi}}&=& \frac{1}{24} \left(3 \varphi ^2 (\kappa +4 \lambda -\mu_\Sigma )+18 \mu_\text{SSI}-24 m_\Sigma^2+2 T^2 (8 \kappa +17 \lambda )\right).
\eea
As a result of the spurion analysis described in the previous section, the masses of the $X$ and the $\pi$ mesons are no longer degenerate in $N_F =4$ case, although the thermal contributions to the spectrum remain degenerate.  
We show an example of a benchmark for the resulting thermal potential in the $N_F=4$ case in Fig.\ref{fig:thermalpotential}. 

\subsubsection{The $\mu_{\Sigma}$ term during the phase transition}\label{sec:instantons}
The determinant interaction $\left( \mu_\Sigma\, { \rm det}\Sigma + h.c. \right)$ in the potential \eqref{eq:linearsigma} is known to be generated by instanton interactions \cite{tHooft:1976snw}. It is therefore reasonable to assume that the term is proportional to the strength of instanton effects~\cite{Pisarski:1983ms}. The temperature scaling of this quantity at large temperatures  (and small gauge couplings) is well approximated by the dilute instanton gas approximation (DGA), as large-scale instantons are suppressed \cite{Gross:1980br}.
However, as $T\rightarrow T_c$ and $\alpha_S \to 1$ (near the confinement scale),   large-scale instantons are no longer suppressed. At low temperatures and strong coupling the instantons can therefore not be considered well-separated, and the DGA is no longer a good approximation. How the instanton density scales during and after the confinement phase transition is currently an open question.

A related, well studied quantity is the topological susceptibility,
\begin{equation}
    \chi(T) = \partial_\theta^2 F(\theta,T) = \int \frac{d\rho}{\rho^5}\, d(\rho,T)
\end{equation}
where $F(\theta,T)$ is the $\theta$-dependent free energy and where $d(\rho,T)$ is the instanton density as a function of instanton size $\rho$ and temperature.
The finite temperature behavior of the topological susceptibility $\chi(T)$ is of interest to the lattice community as well as the axion and dark matter communities, because the misalignment production of light fields depends sensitively on its behavior at finite temperature (see e.g. \cite{Frison:2016vuc,Dine:2017swf}). 
The DGA can be used to compute $\chi(T)$ at high temperatures, which results in $\chi(T) \sim T^{-8}$ (we include a brief review of this calculation in appendix \ref{sec:DGA}). Lattice simulations agree with this prediction above $T \gtrsim 1.5 \,T_c$, and indicate a flattening off of the temperature evolution around $T\sim T_c$ \cite{Frison:2016vuc}. Motivated by these results, we will use $d\mu_\Sigma/dT =0$ for $T_N < T_c$ in the following, but alert the reader that this issue should be revisited if further lattice results become available. A strong temperature dependence of the parameter $\mu_\Sigma$ would mean a larger nucleation rate than calculated below, and an overall decrease of the gravitational wave amplitude.

\subsection{Gravitational wave spectra}
\label{sec:GW-spectra}
In the linear sigma model detailed above, the dynamics of the phase transition are captured by the diagonal field $\varphi$, which has vaccuum expectation value $\varphi=0$ at high temperatures, and $\varphi=f_\Sigma$ after the transition. As is well known, the tunneling between the two vacua of this field is described by the scalar bounce $\varphi_{\rm c}( r,T)$, a
spherically symmetric classical solution to the Euclidean equations of motion \cite{Coleman:1977py}. 
\bea \pdd{\varphi}{ r} + \frac{2}{r} \pd{ \varphi }{ r}   - \pd{V(\varphi,T)}{\varphi} =0 \eea
We use a combination of a shooting algorithm and a finite difference technique to solve this equation at different temperatures, using the 1-loop thermal potential described in the previous section. From this solution, the thermal parameters of the phase transition are derived. 

Firstly, the nucleation temperature $T_N$ is conventionally defined as the temperature for which a particular volume fraction is in the new phase. We will use,
\bea p(t_N) t_N^4 =  \left(\frac{M_p}{T_N}\right)^{4} \left(\frac{45}{16 \pi ^3 g_*}\right)^2  \,e^{-S_E/T_N} = 1 \eea
where $p(t)$ is the nucleation probability per unit time per unit volume, $t_N$ is the nucleation time, and $S_E$ is the Euclidean action evaluated at the bounce solution $\varphi_{\rm c}( r,T_N)$. We have assumed radiation domination to relate $t_N$ and $T_N$. 
Secondly, the nucleation rate is captured by the parameter $\beta$ (conventionally normalized to the Hubble rate), and can also be related to the bounce action,
\begin{equation}
    \frac{\beta}{H} \sim\left. T \, \frac{d (S_E/T)}{dT} \right|_{T=T_N} \ .
\end{equation}
Then, importantly, the latent heat can be defined by,
\begin{equation}
    \alpha = \frac{\mathcal{L}}{\rho_N} \sim \left.\frac{1}{\rho_N} \left(\Delta V - \frac{T}{4} \Delta \dd{V}{T}\right)\right|_{T=T_N} \ ,
\end{equation}
where the symbol $\Delta$ indicates that the quantity is to be evaluated on either side of the wall (with a relative sign), and where $\rho _N= \pi ^2 g^* T_N ^4/30$ is the equilibrium energy density at $T_N$, assuming radiation domination. 

As was also argued in \cite{Helmboldt:2019pan}, it is likely that the chiral phase transition does not exhibit runaway behavior: that is, the bubble walls do not keep accelerating until the bubbles collide. The field $\varphi$ couples to several other bosonic degrees of freedom which will result in friction on the bubble wall. For a non-runaway transition, then, the gravitational wave spectrum resulting from colliding acoustic waves in the plasma is expected to dominate.  
The thermal parameters can then be used to find predictions for this gravitational wave spectrum \cite{Hindmarsh:2017gnf,Weir:2017wfa},\footnote{For strongly supercooled transitions, these spectra have to be modified to reflect that the sound waves do not last longer than a Hubble time \cite{Ellis:2018mja,Ellis:2019oqb}.}
\bea
\Omega _{\rm sw}  h^2  &=& 8.5 \times 10^{-6} \,
\kappa_f^2 \alpha^2 
\,v_w
\left( \frac{100}{g_*} \right)^{1/3} 
\left( \frac{\beta}{H} \right)^{-1}   \times \left( \frac{f}{f_{\rm sw}} \right) ^3 \left( \frac{7}{4+3\left( \frac{f}{f_{\rm sw}}\right) ^2} \right)^{7/2} \\
 f_{\rm sw} &=& 8.9 \left( \frac{z_p}{10} \right) \frac{1}{v_w} \left( \frac{\beta}{H} \right) \left( \frac{T_N}{ 100\, \text{GeV}} \right) \left( \frac{g_* }{100} \right)^{1/6} \, \mu{\rm Hz} \ ,
\eea
Here $z_p$ is a simulation derived factor, which we take to be $z_p=5$ in the following \cite{Hindmarsh:2017gnf}. For $v_w \rightarrow 1$, the efficiency parameter is well approximated by \cite{Espinosa:2010hh},
\begin{equation}
    \kappa _f \sim \frac{\alpha }{0.73+0.083 \sqrt{\alpha } +\alpha}
\end{equation}


\subsubsection{$N_F = 3$}

Here we discuss the thermal parameters of the $N_F = 3$ phase transition and the resulting gravitational wave spectrum.
We will express our results as a function of the ratio of physical masses $m_{\eta'}/m_\varphi$. For $N_F = 3$, this ratio is given in terms of parameters of the linear sigma model by,
 \bea \label{eq:NF3ratio} \frac{m_{\eta'}}{m_\varphi}&=&
 \sqrt{3} \sqrt{\frac{1}{x+1}} < \sqrt{3} \\
 x &=& \frac{4 m_\Sigma^2 (\kappa +3 \lambda )}{\mu_\Sigma  \left(\mu_\Sigma +\sqrt{\mu_\Sigma ^2+4 m_\Sigma^2 (\kappa +3 \lambda )}\right)}.
 \eea
Here the inequality is derived from the mass spectrum \eqref{eq:massesNf3}; realness of the physical masses implies $\kappa + 3 \lambda >0$ and $\mu_\Sigma^2 \geq 0$ in the linear sigma model.\footnote{A weaker constraint comes from \eqref{eq:vevNf3}, $\kappa + 3 \lambda > -\mu_\Sigma^2/4 m_\Sigma^2$.} As we will see below, the upper bound on the ratio $m_{\eta'}/m_\varphi$ has a nontrivial implication for the gravitational wave spectrum from this class of chiral phase transitions.

We perform the bounce calculation described in the previous subsection, with the one-loop thermal potential described in section \ref{sec:thermalV}, for 50 parameter points. We use parameters for which the ratio \eqref{eq:NF3ratio} lies in the range $m_{\eta'}/m_\varphi=[1,\sqrt{3}]$, and the
meson mass limits evade the constraints described in section \ref{sec:masses}: $m_{\eta'}/\text{GeV} = [10^3,2 \times 10^4] $ ; $m_{\pi}/\text{GeV} = [8 \times 10^2, 10^4]$. We use $(m_{X}-m_{\pi})/\text{GeV} = [2 \times 10^2,2 \times 10^4]$, recognizing that $m_{X}>m_{\pi}$ for any choice of parameters, cf. \eqref{eq:massesNf3}. 

\begin{figure}
    \centering
    \includegraphics[width=0.45\textwidth]{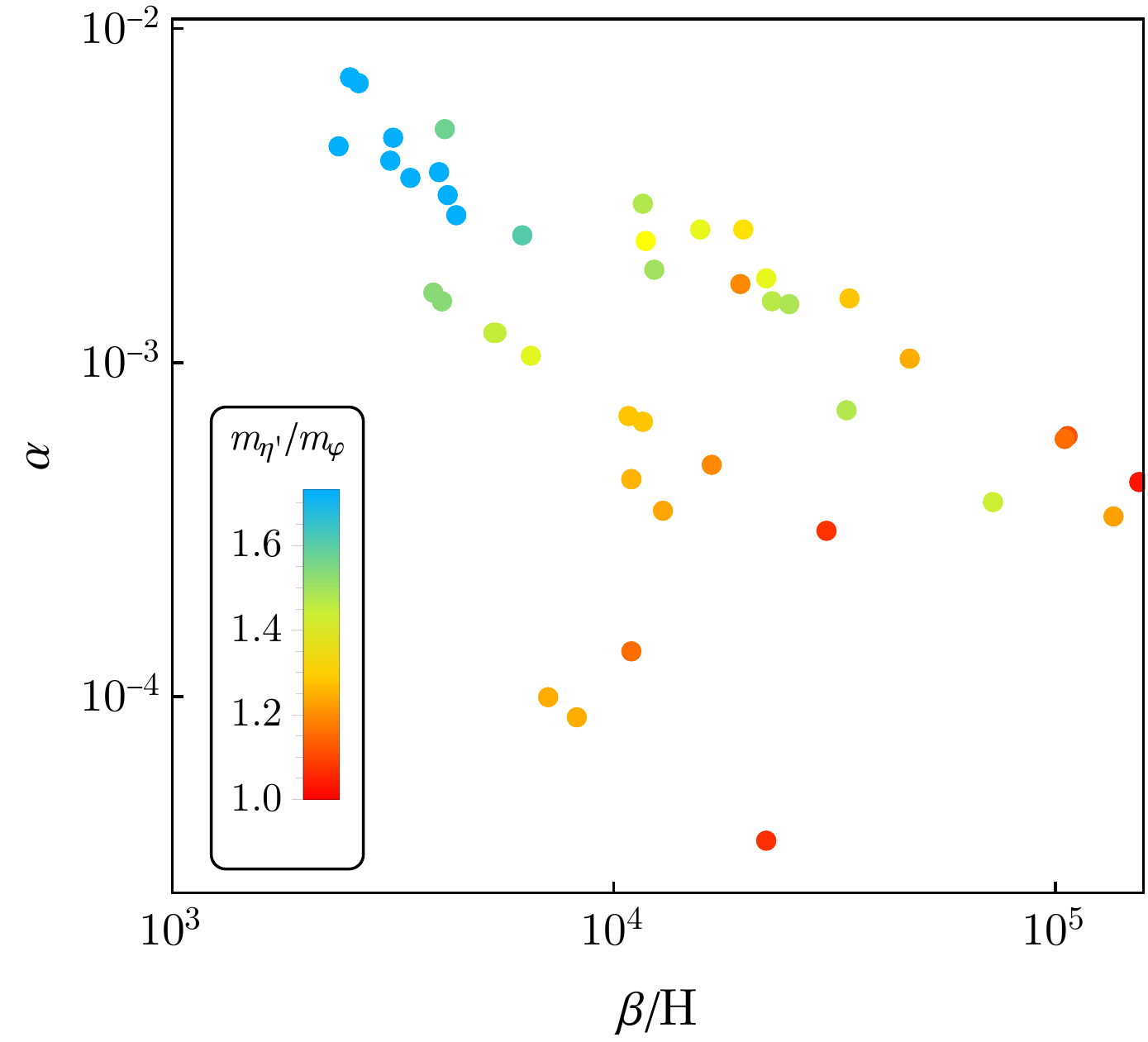}
    \includegraphics[width=0.47\textwidth]{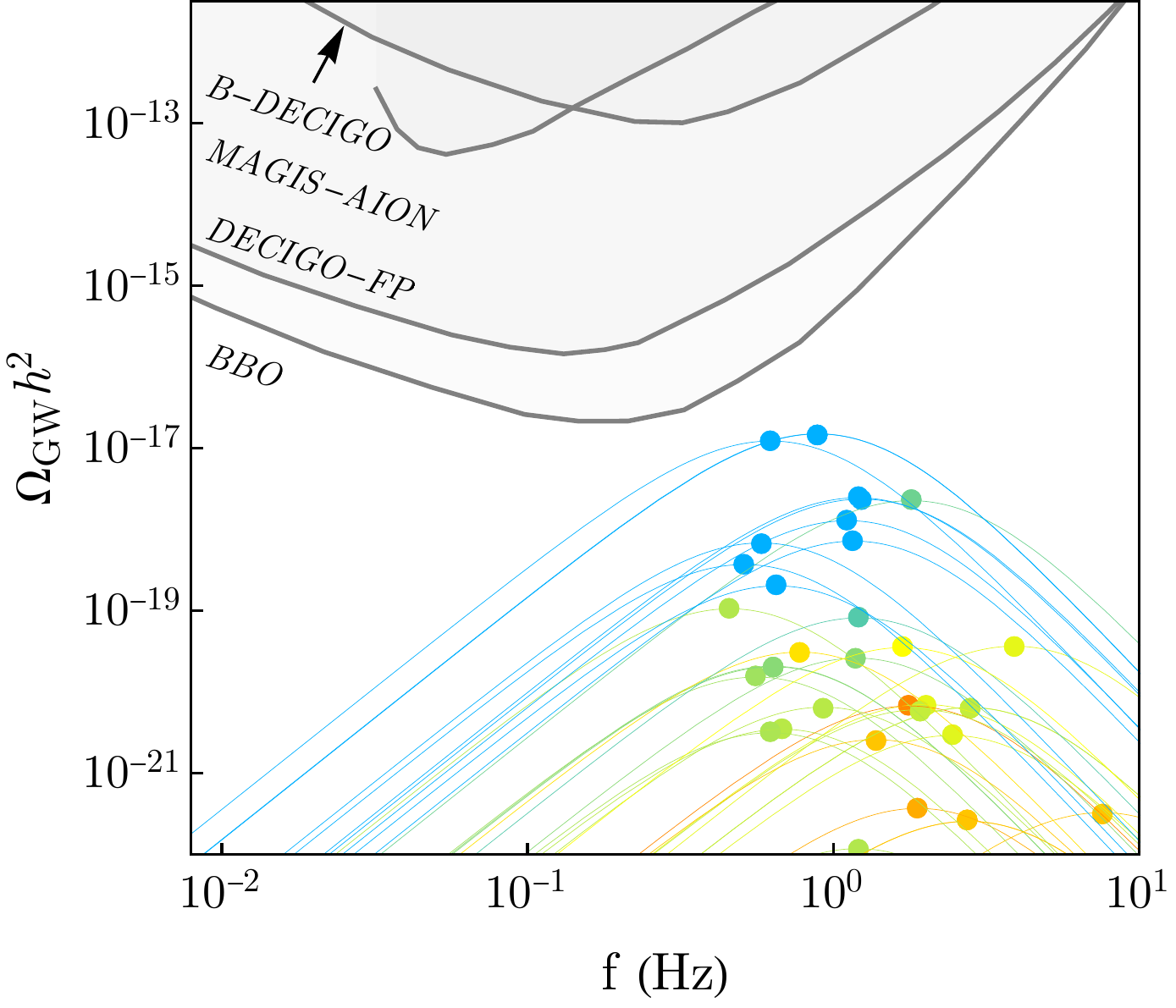}
    \caption{Thermal parameters (left) and gravitational wave signatures (right) of a chiral phase transition with $N_F=3$ and $g_* = 114$, for the benchmarks described in the text. For clarity, the peaks of the spectra are indicated with a point. 
    The experimental projections for the Big Bang Observer (BBO) \cite{Crowder:2005nr,Cutler:2005qq,Harry:2006fi}, Fabry-Perot DECIGO (original proposal) \cite{Yagi:2013du}, B-DECIGO (scaled-down version) \cite{Isoyama:2018rjb}, and MAGIS-AION (space-born) \cite{Graham:2016plp,Graham:2017pmn,AIONtalk} are plotted.
    It is seen that the latent heat parameter $\alpha$ is typically larger, and the rate parameter $\beta/H$ is typically smaller for larger ratio $m_{\eta'}/m_\varphi$, associated with a stronger gravitational wave signature (the panels use the same color scaling).}
    \label{fig:NF3GW}
\end{figure}
 
We plot our results in the Fig.~\ref{fig:NF3GW}, along with projected constraints from various future gravitational wave experiments. The limits plotted here are for power-law spectra, for which the signals in different frequency domains are correlated \cite{Thrane:2013oya}. The most sensitive experiment in this observational window is the Big Bang observer (BBO), a proposed fourth-generation space based interferometer experiment, in a hexagram configuration  \cite{Crowder:2005nr,Cutler:2005qq,Harry:2006fi}. The Deci-Hertz Interferometer Gravitational Wave Observatory (DECIGO) is a space-based Fabry-Perot interferometer which was proposed as early as 2001 \cite{Seto:2001qf,Yagi:2013du}.
B-DECIGO \cite{Isoyama:2018rjb} is the scaled-down version of DECIGO (the B stands for "basic"). Finally, AION and MAGIS are proposed experiments which apply the recently proposed atomic interferometery technique \cite{Graham:2016plp,Graham:2017pmn,AIONtalk}.

It is clear from the right panel of Fig.~\ref{fig:NF3GW} that the gravitational wave predictions from this model will not be probed by any currently proposed gravitational wave experiments. This result can be primarily explained by the small latent heat released in this transition, compared to the large radiation energy density at the time of bubble nucleation. "Dark" QCD models, which do not couple to the Standard Model (and as such evade the experimental constraints described in section \ref{sec:masses}) may feature phase transitions at lower scales. Such models may therefore in principle predict larger gravitational wave amplitudes.

Although the signal is not observable, one can still make a few interesting observations.
From the left panel of Fig.~\ref{fig:NF3GW}, it is seen that the latent heat parameter $\alpha$ and the nucleation rate parameter $\beta$ correlate positively and negatively with the ratio of masses $m_{\eta'}/m_\varphi$ respectively. 
These correlations are also true for the ratio $ f_{\Sigma}/T_c$ (the ratio of the value of the VEV to the temperature at which both phases are degenerate - sometimes referred to as the strength of the transition).
Since the mass of the dynamical axion $m_{\eta'}$ is determined in part by instanton effects as described in the previous subsection, this correlation motivates further study of this parameter at low temperatures.

\subsubsection{$N_F = 4$}
\begin{figure}
    \centering
    \includegraphics[width=0.45\textwidth]{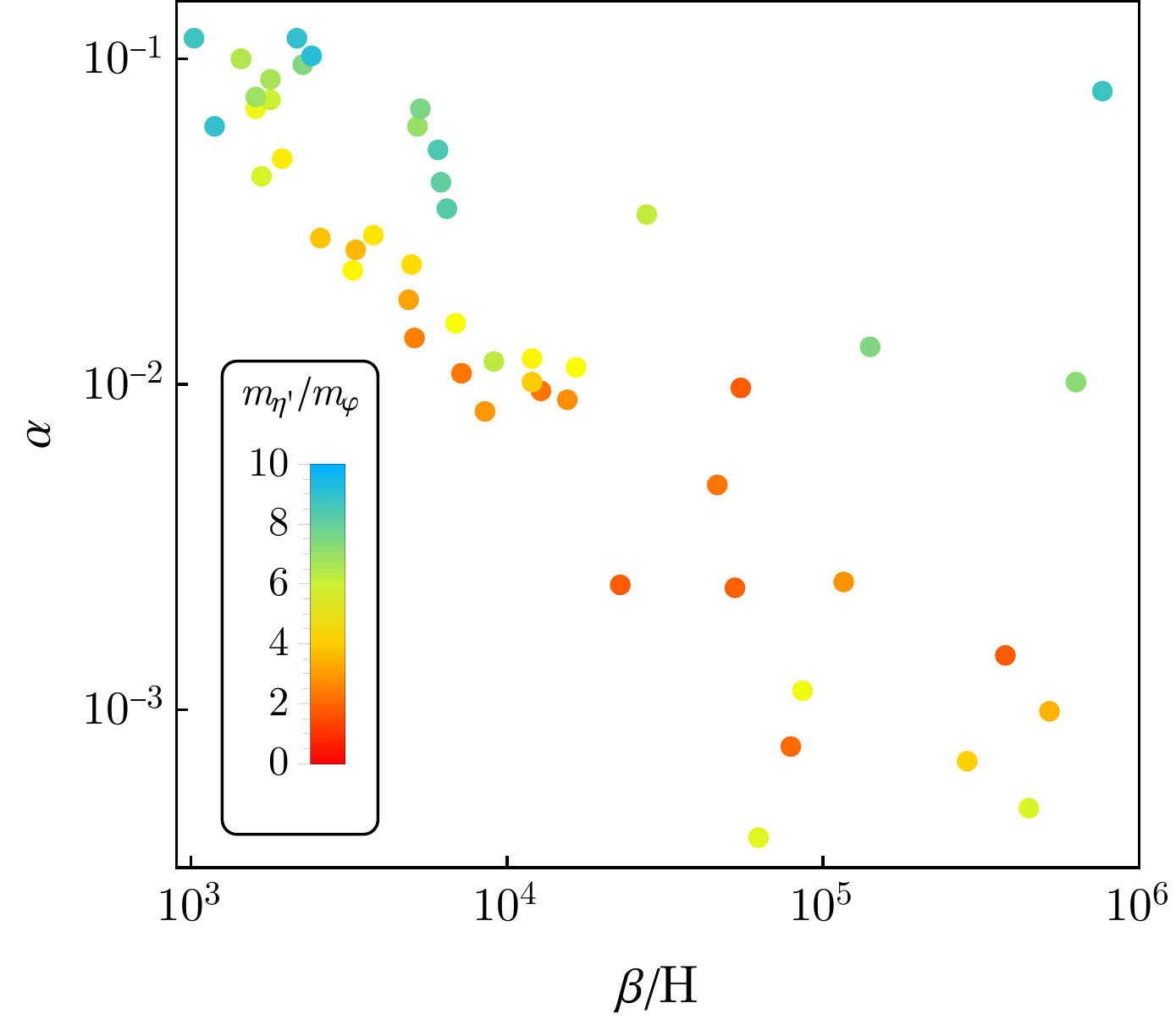}
    \includegraphics[width=0.455\textwidth]{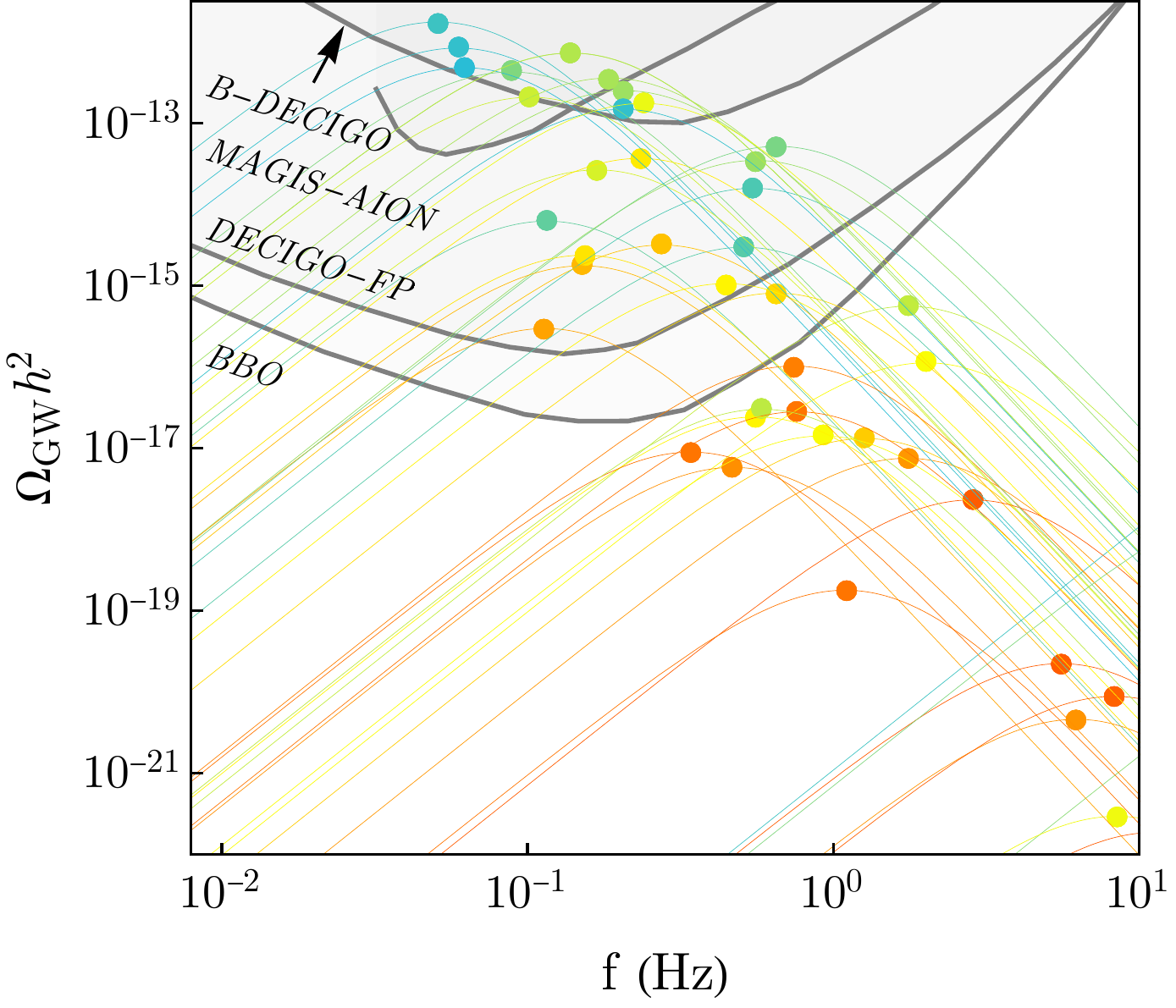}
    \caption{Thermal parameters (left) and gravitational wave signatures (right) of a chiral phase transition with $N_F=4$ and $g_* = 120$, for the benchmarks described in the text. As for $N_F =3$, it is seen that the latent heat parameter $\alpha$ is typically larger, and the rate parameter $\beta/H$ is typically smaller for larger ratio $m_{\eta'}/m_\varphi$, associated with a stronger gravitational wave signature (the left and right panels use the same color scaling).}
    \label{fig:NF4GW}
\end{figure}

For $N_F = 4$, we may again express the ratio of physical (zero temperature) masses in terms of parameters of the linear sigma model,
 \bea \frac{m_{\eta'}}{m_\varphi}= \sqrt{2} \sqrt{\frac{\mu_\Sigma }{\kappa +4 \lambda -\mu_\Sigma }}
 \eea
It is seen that in this case, the ratio is not bounded from above. For $N_F = 4$, we will study a larger range, $m_{\eta'}/m_\varphi=[1,10]$, noting that a large mass ratio is a natural expectation of models such as \cite{Gaillard:2018xgk}. 

We repeat the calculation described in the previous subsection, for 50 parameter points as before. We choose parameters such that the meson masses are in agreement with experimental constraints, 
$m_{\eta'}/\text{GeV} = [10^3,2 \times 10^4] $ ;
$m_{\pi}/\text{GeV} = [8 \times 10^2, 10^4]$; 
$m_{X}/\text{GeV} = [8 \times 10^2, 10^4]$; and
$m_{\eta'_\psi}/\text{GeV} = [8 \times 10^2, 10^4]$.
As explained in section \ref{sec:masses}, the light axion field $\eta'_{\psi}$ is subject to weaker constraints. Here we will use $m_{\eta_\psi'}/\text{GeV} = [1,10^3]$.

We plot the results in Fig.~\ref{fig:NF4GW}. In contrast to the $N_F=3$ case, we note that the predicted gravitational wave spectrum of the $N_F =4$ case may be observed at future interferometer experiments, in line with the naive expectation from the behavior observed in the previous subsection, and the larger ratio $m_{\eta'}/m_\varphi=[1,10]$~\footnote{This $\frac{ m_{\eta'}}{ m_\varphi}$ upper value estimate comes from the fact that (assuming large separations of scales $\mu_{SSI}$ and $\Lambda$) the $\eta'$  gets its mass from confinement and therefore $m_{\eta'}^2 \sim \frac{ \Lambda^4 }{ f^2}$, where $f$ is the chiral symmetry breaking scale and is estimated to be $\Lambda \leq 4\pi f$. Then $m_{\eta'}^2 \leq \left( 4\pi \right)^2 \Lambda^2$ and since $m_\varphi \sim \Lambda$, then $\frac{ m_{\eta'} }{ m_\varphi} \leq 10$.}. In particular, for ratios $m_{\eta'}/m_\varphi \gtrsim 7$, the gravitational wave signal may be detected by atom-interferometers AION and MAGIS. 
From the left plane of Fig.~\ref{fig:NF4GW}, we note that the latent heat released by transitions with larger ratio $m_{\eta'}/m_\varphi$ is larger, while the predicted nucleation rate is smaller. 

The result is interesting, in particular in light of the consideration in subsection \ref{sec:instantons}. The realization of the large mass ratio relies crucially on the value of the parameter $\mu_\Sigma$ at the nucleation temperature, and thus on the explicit breaking of the $U(1)_A$ symmetry. As such, further study of the finite temperature behavior of the instanton density are well-motivated.

\section{Discussion}
This paper has discussed the gravitational wave signatures of models of dynamical axions and confinement at the TeV scale. 
The order of the confinement phase transition relies on the number of light fermions at the confinement scale. An analytic argument based on an expansion in $ \epsilon = 4-d$ \cite{Pisarski:1983ms} implies that phase transitions with $N_F \geq 3$ are first order, and therefore feature a gravitational wave spectrum.

Using the linear sigma model, we studied the cases $N_F = 3$ and $N_F =4$. The gravitational wave predictions of these models are plotted in Figs. \ref{fig:NF3GW} and \ref{fig:NF4GW} respectively. We note that the predictions of the first model evade experimental detection at the presently proposed gravitational wave observatories. For $N_F = 4$, however, the signals may be observable at the Big Bang Observer (BBO) \cite{Crowder:2005nr,Cutler:2005qq,Harry:2006fi}, (B-)DECIGO \cite{Yagi:2013du,Isoyama:2018rjb}, and the MAGIS-AION atom interferometers \cite{Graham:2016plp,Graham:2017pmn,AIONtalk}.

An interesting result is that the amplitude of the gravitational wave spectrum depends on the ratio of the mass of the dynamical axion $m_{\eta'}$, to the mass of the order parameter of the phase transition, the scalar $\varphi$. In particular, the latent heat released in the transition (conventionally captured in the parameter $\alpha$), and the nucleation rate ($\beta/H$) correlate with $m_{\eta'}/m_\varphi $ positively and negatively respectively.
In some models, a large ratio may be a natural prediction \cite{Gaillard:2018xgk}.

The importance of the ratio $m_{\eta'}/m_\varphi$ suggests interesting questions for future research, in particular about the origin of the $\mu_\Sigma$ parameter in the linear sigma model. 
This term constitutes an explicit breaking of the global $U(1)_A$ symmetry. The determinental interaction originates from instanton dynamics, which is known to have a strong temperature dependence at large temperature. The behavior of this parameter at the confinement scale should be investigated further to allow for more detailed studies of the phase transition.

The analysis in this paper leaves open the question of phase transitions in models with $N_F >4$, such as the recently proposed high-scale color confinement model \cite{Ipek:2018lhm}. 
The determinant operator is irrelevant for $N_F >4$, and the strength of the phase transition in the linear sigma model may instead correlate with a further explicit breaking, or the generation of the $\eta'$ mass in such scenarios. The fact that the mass dimension of $\mu_\Sigma$ depends on $N_F$, while the gravitational wave detection prospects also depend strongly on $\mu_\Sigma$ means studying chiral phase transitions for different values $N_F$ may be interesting. 
At $N_F \gg 3$, a conformal window is known to exist, though its exact location is the topic of ongoing research \cite{Frandsen:2010ej,Tuominen:2012qu,Antipin:2017ebo}.

Our study of gravitational wave signatures shows a sizable hierarchy between the scalar $\varphi$ and pseudo-scalar $\eta'$ is preferred for detection. At the same time, these are TeV-range states which can be searched for at colliders. We used current limits on the colored states to guide our parameter search, but these colored states would also induce loop-level contributions to couplings of the $\varphi$ and $\eta'$ to gluons and hence could be looked for at the LHC via dijet signatures~\cite{ATLAS:2015nsi,Sirunyan:2018xlo}.  One can envision a future situation where a signature in gravitational waves is found, and that would guide searches for two correlated states in dijets at colliders. Further angular analysis of the dijet final state could also allow us to determine the CP properties of these states, and robustly support the origin of the gravitational wave signature as manifestation of a dynamical axion explanation of the QCD CP problem. This connection between gravitational wave signatures and collider searches is an area that we plan to develop further.

{\bf Note added. }\emph{While this paper was in its final stages, \cite{Helmboldt:2019pan} appeared on the arXiv. Although the focus of the current work is different, the analysis overlaps partially with the analysis in \cite{Helmboldt:2019pan}. To this extent, the works are qualitatively consistent, though different benchmarks and more recent lattice results were used here.}

\section*{Acknowledgement}
The authors thank Belen Gavela for useful discussions. DC further thanks David Morrissey, Andrew Long, Tim Tait, Seyda Ipek, and Graham White for discussions, and RH thanks Pablo Quilez for discussions. RH also thanks the High Energy Theory Group at Harvard University for their kind hospitality. TRIUMF receives federal funding via a contribution agreement with the National Research Council of Canada and the Natural Science and Engineering Research Council of Canada. RH acknowledges support from the the Spanish Research Agency (Agencia Estatal de Investigaci\'on) through the grant IFT Centro de Excelencia Severo Ochoa SEV-2016-0597. This project has received funding from the European Union's Horizon 2020 research and innovation programme under the Marie Sklodowska-Curie grant agreement No 674896 (ITN Elusives). The work of VS is funded by the Science Technology and Facilities Council (STFC) under grant number ST/P000819/1.

\appendix
\section{Dilute instanton gas approximation}
\label{sec:DGA}
This appendix is meant as a brief review of the dilute instanton approximation (DGA) at finite temperature. In the DGA, the $\theta$-dependent free energy is given by
\begin{equation}
    F(\theta,T) = - \int \frac{d\rho}{\rho^5}\, d(\rho,T) e^{i \theta}
\end{equation}
where $d(\rho,T)$ is the dimensionless instanton density ($\rho$ is the instanton size). The finite temperature behavior of the density is given by \cite{Gross:1980br},
\begin{equation}\label{eq:instantondensity}
    d(\rho,T) = d(\rho,0) \,\exp \left(-12 A(\lambda ) \left(\frac{N_c-N_F}{6}+1\right)-\frac{1}{3} \lambda ^2 (2 N_c+N_F)\right)
\end{equation}
with $\lambda = \pi  \rho  T$  and $$A(\lambda) = c_1 \left(\frac{1}{c_2 \lambda ^{-3/2}+1}\right)^8-\frac{1}{12} \log \left(\left(\frac{\lambda }{3}\right)^2+1\right).$$ 
Here $d(\rho,0)$ is the (dimensionless) zero temperature instanton density - however, this quantity depends on the renormalization scale $\mu$ which may be set equal to $1/\rho$ or to $T$.
The topological susceptibility in the DGA is found from
\begin{equation}
    \chi(T) = \partial_\theta^2 F(\theta,T) = \int \frac{d\rho}{\rho^5}\, d(\rho,T)
\end{equation} 
For $N_F = 3$, the quantity we are interested in may be approximated by, 
\begin{equation}\label{eq:DGAmuSigma}
    \mu_\Sigma \sim f^6\int d\rho\,\rho^4 d(\rho,T)
\end{equation}
using dimensional analysis and $\Sigma \sim \bar{q}q/f^2$.

\begin{figure}
    \centering
    \includegraphics[width=0.45\textwidth]{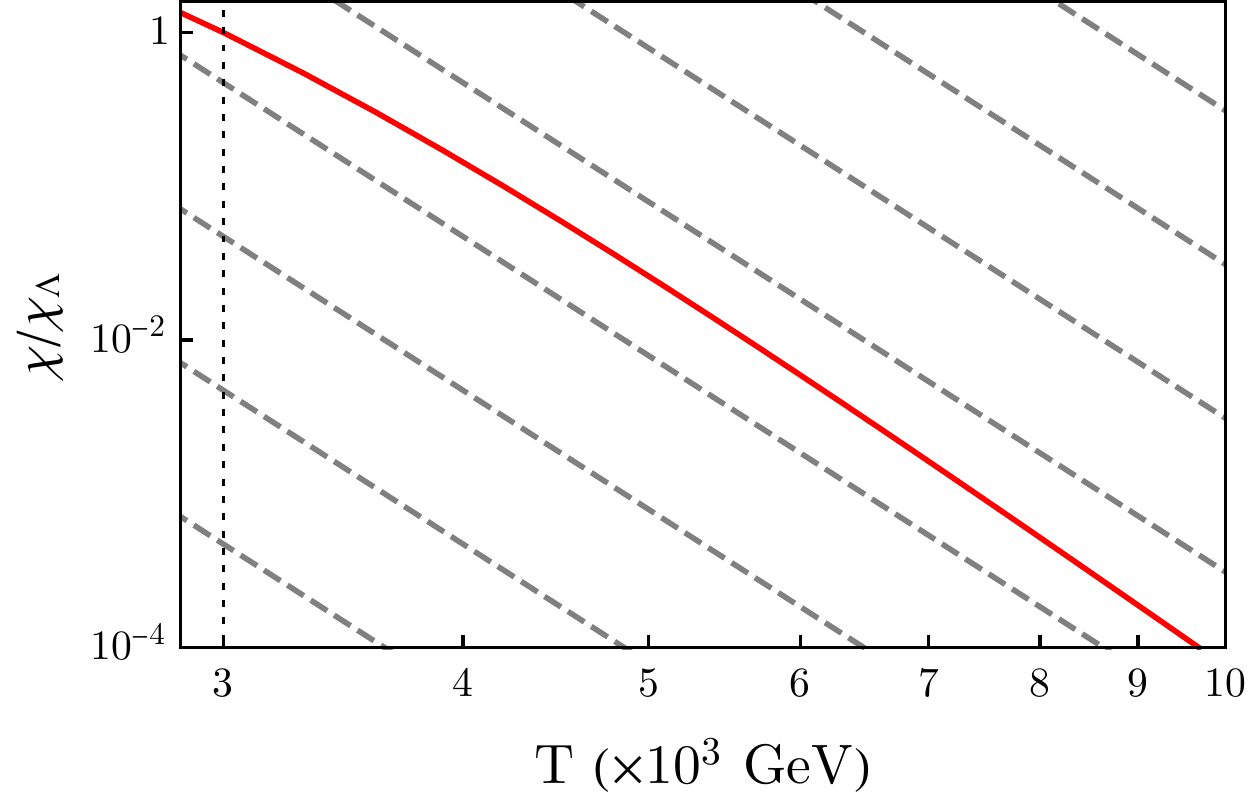}
    \includegraphics[width=0.45\textwidth]{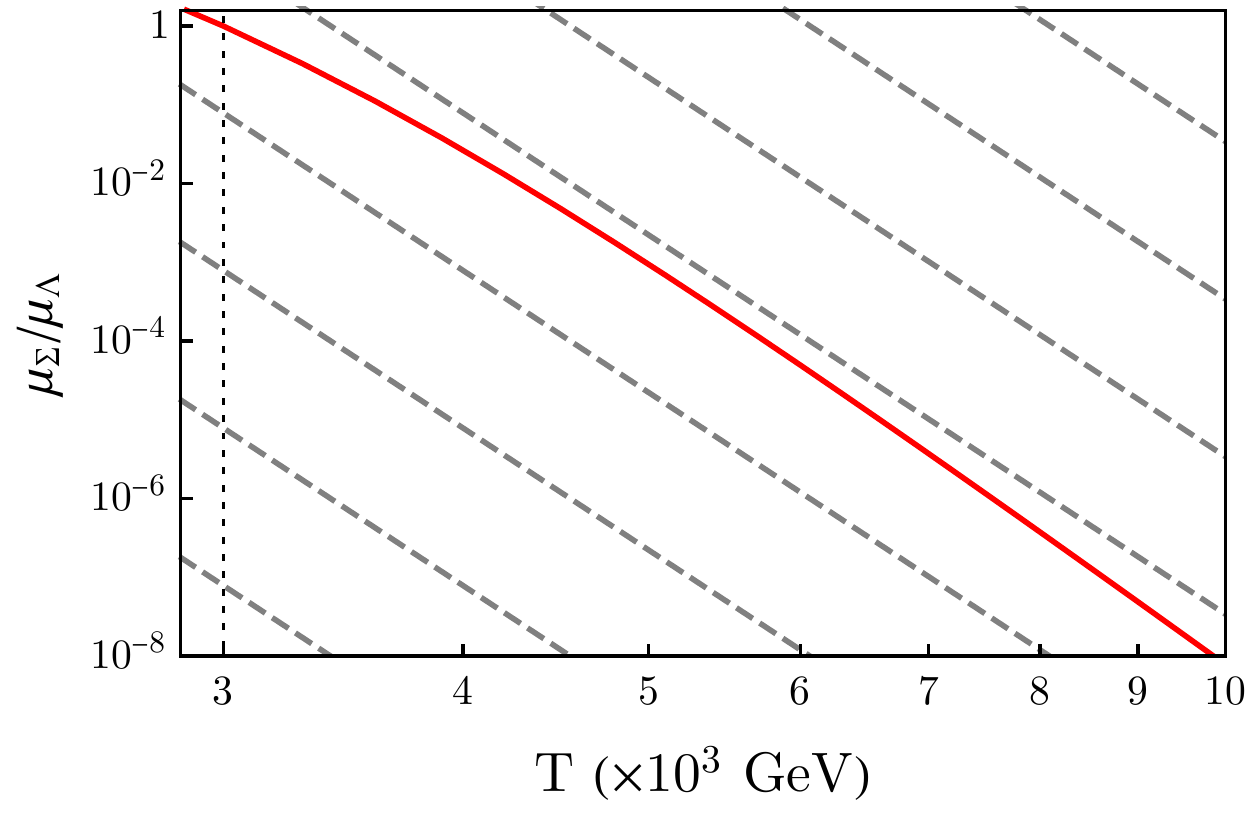}
    \caption{Topological susceptibility in the DGA for $N_F =3$ and $\Lambda =3$ TeV confinement scale. The dashed gray lines are plotted for reference only, and have slope $T^{-8}$ (left) and $T^{-16}$ (right).}
    \label{fig:DGA_ps}
\end{figure}

The topological susceptibility and \eqref{eq:DGAmuSigma} in the DGA approximation are plotted in Fig. \ref{fig:DGA_ps}.
It is seen that for high temperatures, the slope approximates $T^{-8}$ as is known. However, this approximation is no longer trusted in the regime in which the PT takes place. Lattice studies in the regime $T< T_c$ are inconclusive, though it is generally expected that at very low temperatures $\chi(T) \sim T^{-n}$ where $0\lesssim n\ll8$. 

\bibliographystyle{JHEP}
\bibliography{references}
\end{document}